# Optical diffraction tomography using a digital micromirror device for stable measurements of 4-D refractive index tomography of cells

Seungwoo Shin [a*], Kyoohyun Kim [a*], Taeho Kim [b*], Jonghee Yoon [a], Kihyun Hong [c†], Jinah Park [b†], and YongKeun Park [a,c†]

[a]Department of Physics, Korea Advanced Institute of Science and Technology, Daejeon 34141, Republic of Korea;
E-mail: yk.park@kaist.ac.kr

[b]School of Computing, Korea Advanced Institute of Science and Technology, Daejeon 34141, Republic of Korea;
E-mail: jinahpark@kaist.ac.kr

[c]TOMOCUBE, Daejeon 34051, Republic of Korea; E-mail: khhong@tomocube.com

[*]These authors contributed equally to this work.

**ABSTRACT**

Optical diffraction tomography (ODT) is an interferometric microscopy technique capable of measuring 3-D refractive index (RI) distribution of transparent samples. Multiple 2-D holograms of a sample illuminated with various angles are measured, from which 3-D RI map of the sample is reconstructed via the diffraction theory. ODT has been proved as a powerful tool for the study of biological cells, due to its non-invasiveness, label-free and quantitative imaging capability. Recently, our group has demonstrated that a digital micromirror device (DMD) can be exploited for fast and precise control of illumination beams for ODT. In this work, we systematically study the precision and stability of the ODT system equipped with a DMD and present measurements of 3-D and 4-D RI maps of various types of live cells including human red blood cells, white blood cells, hepatocytes, and HeLa cells. Furthermore, we also demonstrate the effective visualization of 3-D RI maps of live cells utilizing the measured information about the values and gradient of RI tomograms.

## 1. INTRODUCTION

Optical diffraction tomography (ODT) is an interferometric microscopy technique that enables to investigate three-dimensional (3-D) refractive index (RI) distribution of transparent samples such as biological cells and tissues [1-3]. ODT was first theoretically established by E. Wolf [4], and then experimentally demonstrated from the late 1970s [5, 6]. Recently, ODT has regained wide interest and grown rapidly particularly in the field of biophotonics [1, 2], because it provides unique advantages of non-invasiveness, label-free and quantitative imaging capability over existing 3-D optical microscopy. Several approaches for ODT techniques have been developed [7-11] and exploited in various applications in life science and medicine including hematology [12-15], parasitology [16, 17], phytoplankton[18], and microbiology [19, 20], as well as industrial field such as downy hairs [21] and optical plastic lenses [22].

Essentially, ODT solves an inverse scattering problem; 3-D RI map of a weakly scattering semi-transparent object is reconstructed from measurements of multiple 2-D holograms of the object illuminated with various angles [4, 23-25]. In order to systemically control the illumination beam, a galvanometric mirror has been typically utilized [10, 26]. Although the use of a galvanometric mirror in ODT is straightforward and enables fast control of illumination angles, it suffers from several practical limitations: jittering and positioning errors resulted from electric noise and nonlinear response at high voltage, and expensive unit. Furthermore, a dual-axis galvanometer cannot be conjugated accurately due to the geometry of the galvanometer which makes undesired quadratic phase distortion on the illumination beam.

Recently, in order to overcome these limitations of mechanical rotating mirrors, a liquid crystal spatial light modulator (SLM) has been used to control the angles of the illumination beam [27]. In spite of the high stability and the wavefront correction ability of the use of an SLM, the intrinsically slow response of liquid crystal and expensive costs of the SLM are technical limitations. More recently, our group presented a method to utilize a digital micromirror device (DMD) for ODT [17]. A DMD contains millions of individually switchable micromirrors which enables intensity modulation of incident beams at the speed of a few to tens of kHz. For controlling the illumination angle for ODT using the DMD, the principle of Lee hologram is employed [28]. A series of binary amplitude gratings patterns are displayed onto the DMD, and the first diffracted beam is spatially filtered so that it can be used active illuminations. The spatial period of the projected grating pattern determines the spatial frequency of the first diffracted beam, resulting the precise control of angle of the illumination beam impinging onto a sample. Two-axis angle control is also straight forward with a DMD, because grating periods over x- and y- directions can be independently assigned. In addition, general wavefront correction is also enabled using the DMD, because arbitrary phase delay can be applied using the Lee hologram principle. Furthermore, thanks to the mass production, the price of a DMD is about a few hundred USD.

Here, we systematically study the precision and stability of the ODT system equipped with a DMD and also present measurements of 3-D and 4-D RI map of various types of live cells including human red blood cells (RBCs), white blood cells (WBCs), hepatocytes, and HeLa cells. By optimizing the beam path and making a compact optical configuration, a user-friendly 3-D holographic microscope is produced. In addition, the tomogram reconstruction speed is further enhanced by just-in-time (JIT) compilation, CUDA computing, and customized algorithm optimization.

## 2. METHODS

### 2.1 Optical setup

The optical setup for ODT using a DMD is presented in Fig. 1. To control the angles of the illumination beam, a series of the Lee holograms [29] corresponding plane waves with various spatial frequencies are displayed onto the DMD. Then, the first order diffracted beam from the DMD is filtered and projected onto a sample plane. The beam which passes the sample is further projected onto the CCD plane, where it interferes with a reference beam that is split from a fiber-type beam splitter. The resultant spatially modulated holograms are recorded by the CCD, from which 2-D complex optical fields containing both the amplitude and phase image of the sample are retrieved via the field retrieval algorithm [30]. The optical system is fully integrated into a stand-alone instrument, as shown in Fig. 1b, and the instrument is commercially available via TOMOCUBE.

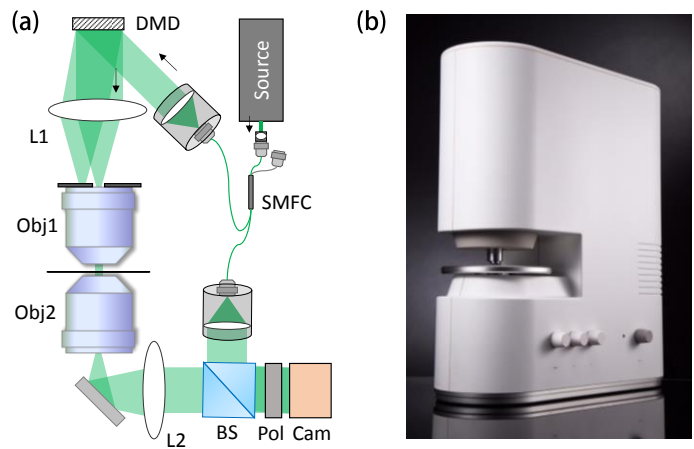

**Fig 1.** (a) Optical setup for optical diffraction tomography using a digital micromirror device. The laser beam is divided by a 2 × 2 single-mode fiber coupler (SMFC) into the sample and reference arms in Mach-Zehnder interferometry; L1–L2, lenses ($f_1$ = 250 mm; $f_2$ = 180 mm), Obj1-Obj2, objective lenses (UPLSAPO 60XW, 60×, NA = 1.2, Olympus Inc., Japan). (b) Commercialized version of the 3-D holographic microscope.

In order to reconstruct a 3-D RI map of a sample, typically 101 holograms are recorded, which correspond to 100 plane wave illuminations scanned in a circular pattern with the angle of 49° in a medium and one normal illumination.

Using the measured multiple 2-D holograms, a 3-D RI map is reconstructed using the diffraction tomography algorithm with the Rytov approximation. The numerical apertures of objective lenses inevitably limit the angle of acceptance, causing the so-called the missing cone problem. In order to fill up this missing information, non-negativity iterative algorithm is employed. The detailed information about the experimental setup, the MatLab code for the reconstruction algorithm, and the algorithms for solving the missing cone problem are described elsewhere [16, 17, 31].

**2.2 Stability of the illumination control unit using a DMD**

In order to demonstrate the stability and the quality of the illumination control of the system using a DMD, we compared with a system using a dual-axis galvanometric mirror (Fig. 2). For the same interferometric microscopy system, we prepared two independent illumination schemes: one with a DMD and the other with a dual-axis galvanometric mirror (Fig. 2a). The galvanometric mirror was located at the conjugated plane of the DMD, and tilting of the galvanometric mirror controls the angle of illumination.

To test whether the DMD control the illumination angle, we systematically compared the diffracted beams from a sample, a polystyrene microsphere with the diameter of 10 μm, between illuminations with the DMD and the galvanometric mirror (Fig. 2b). For 10 different angles of illumination, the measured phase images for these two illumination schemes shows consistent results. The cross-correlation coefficients of phase images are greater than 0.98 for all the cases. This result clearly indicates that the quality of the illumination control with the DMD is compatible with the case with the galvanometric mirror.

Next, in order to demonstrate the repeatability of the illumination control using the DMD, we repeatedly measured phase images of the polystyrene bead 100 times with 10 different illumination angles per each time. To quantitatively analyze the repeatability of the illumination control, we calculated the cross-correlation coefficient of the measured phase images in 100 repetitions with the first acquisition for each incident angle. As shown in Fig. 2(c), the cross-correlation coefficient values are between 0.99 and 1.0 independent of the number of repetitions and incident angles. This result demonstrates high repeatability of the illumination control with the DMD.

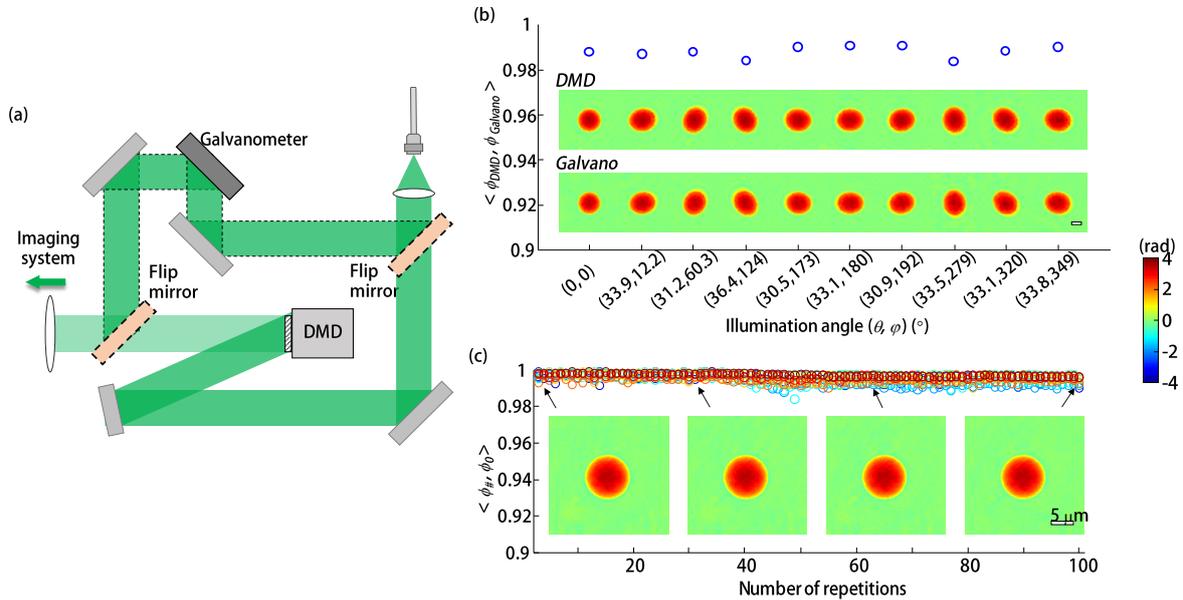

**Fig 2.** (a) Experimental setup to validate the stability of the illumination with a DMD in comparison to the use of a galvanometric mirror. Flip mirrors are used to share the same illumination source and the imaging system for a systematic comparison. Cross-correlation coefficient of 2-D phase images (b) between the same illumination angles controlled by the DMD and the galvanometer and (c) between various number of repetitions

and the initial control. Each incident angle is marked with different color. All of the obtained cross-correlation values are greater than 0.98. The scale bars indicate 5 μm.

## 3. RESULTS

In order to demonstrate the capability of the present setup, we measured and reconstructed 3-D RI distributions of various biological samples including human RBCs, mice WBCs, HeLa cells and hepatocytes (Huh-7). Human RBCs are extracted from healthy donors, and diluted 20 times in Dulbecco's phosphate buffered saline (DPBS, Welgene Inc., Republic of Korea) in order to image individual cells in 20 μm × 20 μm field of view of the detector. RBCs are plated on a cover glass and sandwiched using another cover glass to prevent the sample drying. Mice WBC were collected from mice peritoneal cavity according to a sample preparation protocol [14]. In brief, 7-weeks-old Balb/c mouse (Orient Bio Inc., Republic of Korea) was euthanized with $CO_2$ gas. Then, 5 mL of ice-cold DPBS with 5% FBS was injected into the peritoneal cavity of the mice using a syringe with a 25G needle. Then, injected fluid was collected using a syringe with a 26G needle and centrifuged at 250 g for 8 min. The macrophage and lymphocytes were maintained in RPIM 1640 (Welgene Inc., Republic of Korea) supplemented with 10% heat-inactivated fetal bovine serum (FBS), 1,000 U/L penicillin, and 1,000 μg/mL streptomycin. The prepared WBCs were plated on a 24 × 40 mm cover glass (Marienfeld-Superior Inc., Germany) and sandwiched using another cover glass to prevent the sample drying. HeLa cells and Huh-7 cells were maintained in high glucose Dulbecco's Modified Eagle's Medium (DMEM, LM001-05, Welgene Inc., Republic of Korea) supplemented with 10% heat-inactivated FBS, 1,000 U/L penicillin, and 1,000 μg/mL streptomycin in a humidified incubator at 37 °C and 5% $CO_2$. Cells were plated on a 35 $mm^2$ culture dish (ibidi Inc., Germany) and maintained for 4 - 8 hours in an incubator. The sample preparation protocol was approved by the Institutional Review Board (IRB project number: KA-2015-03).

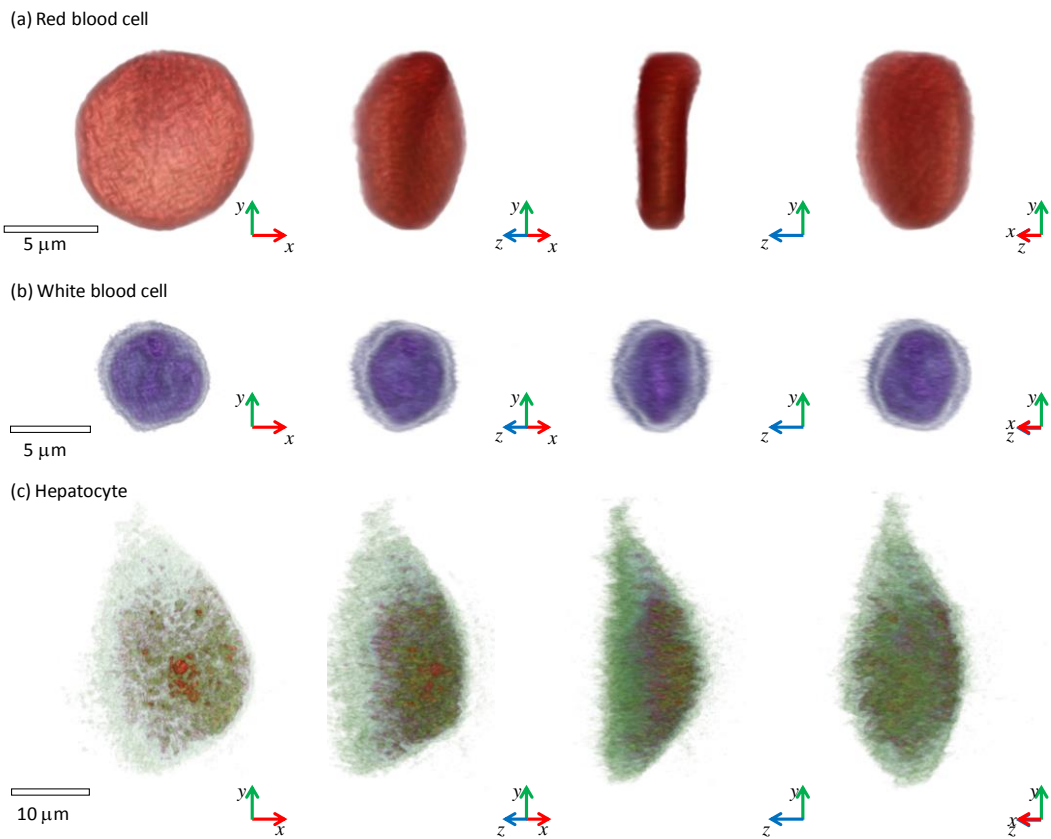

**Fig 3.** Rendered tomograms of various biological cells: (a) Red blood cell, (b) white blood cell, and (c) hepatocyte. Colored arrows indicate the coordinates.

For effective visualization of 3-D RI maps of biological cells, we applied volume rendering with customized transfer functions, which convert the absolute and the local gradient value of RIs into appropriate pseudo-color maps with transparency, for RBCs, WBCs, and eukaryote. The results are shown in Fig. 3. A reconstructed tomogram of a RBC clearly shows distinctive dimple shape in the center region of the cell [Fig. 3(a)]. The reconstructed 3-D RI distribution of a WBC (small lymphocyte) reveals intracellular structures including a nucleus, which resembles an image taken by light microscopy with Giemsa staining, a gold-standard method in clinical settings [Fig. 3(b)]. The reconstructed tomogram of a hepatocyte provides 3-D spatial distribution of intracellular vesicles, presumably lipid droplets [Fig. 3(c)].

In order to demonstrate the applicability of the present setup for biomedical studies, we measured long-time measurements of time-lapse 3-D RI map of eukaryotic cells under cell death. 3-D RI map of individual HeLa cells and hepatocytes under apoptosis and necrosis are sequentially measured for 14 hours in every 30 seconds. As shown in Fig. 4(a), a HeLa cell first exhibits apoptotic blebbing, which is one of the distinctive features of apoptosis. The HeLa cell then detaches from a coverglass and forms a round shape. After 10 hours, overall RI values of the cell became larger since the protein concentration inside the cell increases. The cell is terminated following the encapsulation of intracellular components (Fig. 4a, the right panel). In contrast, necrosis in a hepatocyte exhibits different morphological features as shown in Fig. 4(b). In necrosis, the cell is suddenly collapsed after forming a void region inside the cell (Fig. 4b, the center panel). The release of intracellular components, followed by necrosis – ordered cellular explosion, may induce inflammation processes in neighboring cells. Compared to the recent work employing 2-D optical phase images [32], the present approach directly visualizes the 3-D RI maps during the cell death, allowing more detailed analysis and investigation. These results clearly demonstrate that the present method provides high spatial and temporal resolution so that it can clearly visualize 4-D images of cell dynamics without using labelling agents.

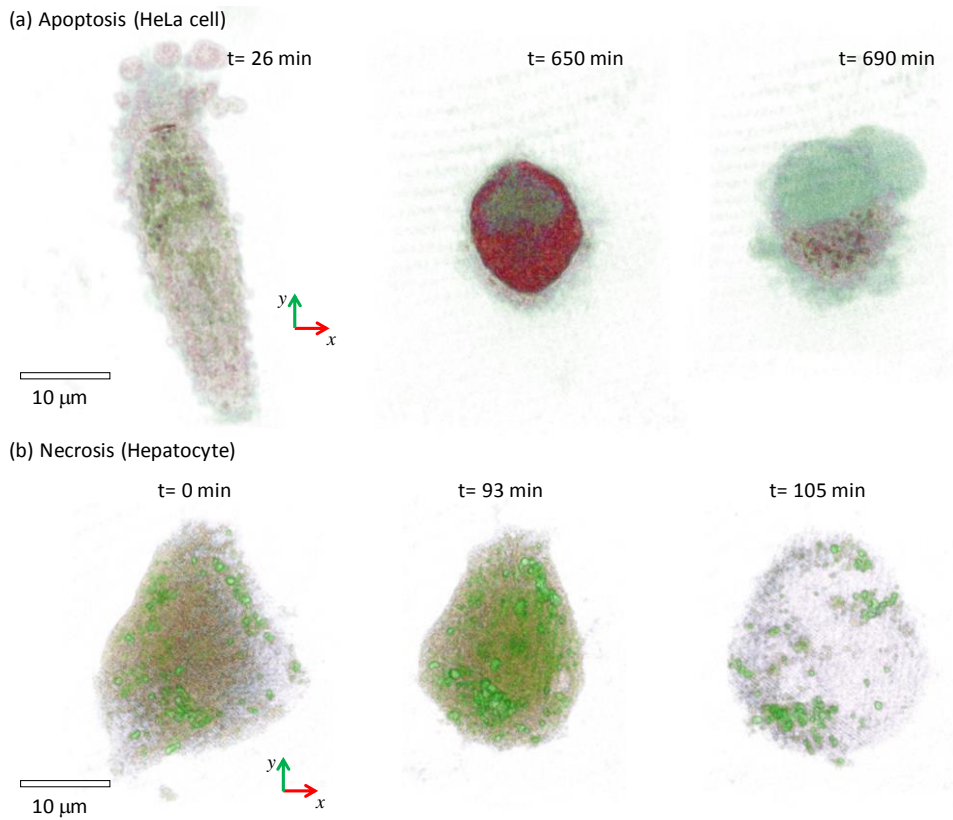

**Fig. 4.** Time-lapse 3-D RI distribution of (a) a HeLa cell, and (b) a hepatocyte suffering apoptosis and necrosis, respectively.

Field retrieval and tomogram reconstruction was performed in a desktop computer. Detailed algorithms can be found elsewhere [16, 31, 32]. Compared to our recent work [28], the tomogram reconstruction speed has been further enhanced by employing just-in-time compilation, CUDA computing, and customized algorithm optimization. The total computation time for reconstructing a tomogram with the size of 192 × 192 × 192 voxels from 101 illumination angles takes 1.16 sec, which is 8.33 times faster than the previous method [28].

## 4. CONCLUSION

In this paper, we report that the ODT system equipped with a DMD is a powerful tool to investigate the structures and dynamics of biological cells. We systematically validate the precision and stability of the ODT system equipped with a DMD. In particular, the direct comparison with the illumination with a dual-axis galvanometric mirror presents the high stability and repeatability of the ODT system with a DMD. Furthermore, we also demonstrate measurements of 3-D and 4-D RI maps of various types of live cells including human red blood cells, white blood cells, hepatocytes, and HeLa cells. The high spatial and temporal resolution of the present method enables the precise measurements of structures and dynamics of individual biological cells with the unique advantageous of 3-D quantitative phase imaging: label-free and quantitative imaging capability. In addition, we also demonstrate the effective visualization of 3-D RI maps of live cells utilizing customized transfer functions; the measured RI values and gradients were effectively transformed into pseudo color and transparently for deter visualization. With these capability and advantages, we believe the ODT with a DMD will be a power and effective method for the investigation of cells and tissues.

Although the current work demonstrates the measurements of 3-D RI map of individual cells, the present approach is sufficiently broad and general, and it will directly offer powerful approaches for various investigation. For example, the optical instrument can be exploited for other applications, including the retrieving angle-resolved light scattering spectra via Fourier Light Scattering Spectroscopy [33-36], synthetic aperture digital holography [37], and measuring phase images of tissue slices [38]. Furthermore, the combination with other optical modality, it may also possible to address hyperspectral quantitative phase imaging [39-41], optical tweezers [42], and the simultaneous measurements of fluorescence and quantitative phase imaging [43, 44]. We anticipate the present simplified method for ODT enables the employment of ODT to various research fields, from hematology [45], pathology [38, 46], neuroscience [47], biophysics [48], reproductive health [49], infectious diseases [50-52], genetic diseases [53, 54], drug efficacy [55, 56], and nanotechnology [57, 58].

## 5. ACKNOWLEDGEMENTS

This work was supported by KAIST, and the National Research Foundation of Korea (2015R1A3A2066550, 2014K1A3A1A09063027, 2012-M3C1A1-048860, 2014M3C1A3052537) and Innopolis foundation (A2015DD126).